\documentstyle[aps,epsfig,psfig]{revtex}
\newcommand{\beq}{\begin{equation}}
\newcommand{\eeq}{\end{equation}}
\newcommand{\beqa}{\begin{eqnarray}}
\newcommand{\eeqa}{\end{eqnarray}}
\newcommand{\half}{\frac{1}{2}}
\newcommand{\gsim}{\buildrel > \over {_\sim}}
\newcommand{\lsim}{\buildrel < \over {_\sim}}

\newcommand{\order}[1]{${\cal O}(#1)$}

\newcommand{\ie}{{\it i.e.}}
\newcommand{\eg}{{\it e.g.}}
\newcommand{\cf}{{\it cf.\ }}

\newcommand{\gev}{{\rm GeV}}
\newcommand{\eq}[1]{Eq.~(\ref{#1})}
\newcommand{\as}{\alpha_s}
\newcommand{\lqcd}{\Lambda_{QCD}}

\newcommand{\PL}[3]{Phys.\ Lett.\ {{\bf#1}} ({#3}) {#2}}

\newcounter{saveeqn}
\newcounter{App} 

\begin{document}
\hbox to\hsize{\normalsize\hfil\rm NORDITA-2000/20 HE}
\hbox to\hsize{\normalsize\hfil\rm LU TP 00-09}
\hbox to\hsize{\normalsize\hfil\rm DAPNIA/SPHN-00-11}
\hbox to\hsize{\normalsize\hfil hep-ph/0003257}
\hbox to\hsize{\normalsize\hfil \protect\today}

\vspace{0.2cm}

\thispagestyle{empty}
\begin{center}
{\Large\bf Semi-Exclusive Production of Photons at HERA}
\vskip .5truecm

{\large\bf Paul~Hoyer}

{\it NORDITA \\Blegdamsvej 17, 2100 Copenhagen, Denmark}

\vskip .5true cm
{\large\bf Martin~Maul}

{\it Department of Theoretical Physics, Lund University \\
S\"olvegatan 14 a, S-223 62 Lund, Sweden}

\vskip .5true cm
{\large \bf Andreas~Metz}

{\it DAPNIA/SPhN, CEA-Saclay \\
F-91191 Gif-sur-Yvette, France}

\end{center}
\begin{abstract}
We study the feasibility of measuring semi-exclusive photon production
$\gamma + p \to \gamma + Y$
at HERA. The cross section of photons produced at large transverse momenta,
recoiling off an inclusive system $Y$ of limited mass, can without photon
isolation cuts be simply expressed in terms of hard PQCD subprocesses and
standard target parton distributions. With the help of event generators we
identify the kinematic region where quark and gluon fragmentation processes can
be neglected. The cross section in this semi-exclusive region is large enough
to be measured with an upgraded HERA luminosity of
${\cal L} = 100\;{\rm pb}^{-1}$.

\noindent
The subprocesses of lowest order in $\alpha_s$ are suppressed at low recoil
masses $M_Y$, compared to higher order gluon exchange,
i.e. BFKL contributions. The
distinct $M_Y$-dependence makes it possible to determine experimentally the
kinematic range where the higher order processes dominate.


\end{abstract}


\newpage

\section{Introduction} \label{secI}
Semi-exclusive reactions of the type $A+B \to C+Y$, where particle $C$
emerges with large transverse momentum and the mass of the inclusive system $Y$
satisfies $\lqcd \ll M_Y \ll E_{cm}$, provide a new tool for probing the
structure of hadrons\cite{Brodsky:1998sr}.
It is analogous to DIS, $ep \to eX$, in the sense that particles $A$ and $C$
form an effective current which probes the target $B$ at large momentum
transfer $t=(p_A-p_C)^2$, producing the inclusive system $Y$.
Semi-exclusive cross sections factorize into a calculable short distance
interaction times a standard target parton $(b)$ distribution $f_{b/B}(x_S,-t)$
according to
\beq
\frac{d\sigma}{dt\,dx_S}(A+B\to C+Y)
= \sum_{b} f_{b/B}(x_S,-t) \frac{d\sigma}{dt} (A b \to C d) \,,
\label{gencross}
\eeq
where
\beq
x_s = -t/(M_Y^2-t)  \label{xsdef}
\eeq
and the momentum transfer $-t$ serves as factorization scale.
For large $-t$ only compact configurations of
particles $A$ and $C$ participate in the reaction, and their re-interactions
with the target are suppressed.
The final parton(s) $d$ merges with the target spectators to form the
inclusive system $Y$ (Fig. 1a).

\begin{figure}[htb]
\center\leavevmode
\epsfxsize=14cm
\epsfbox{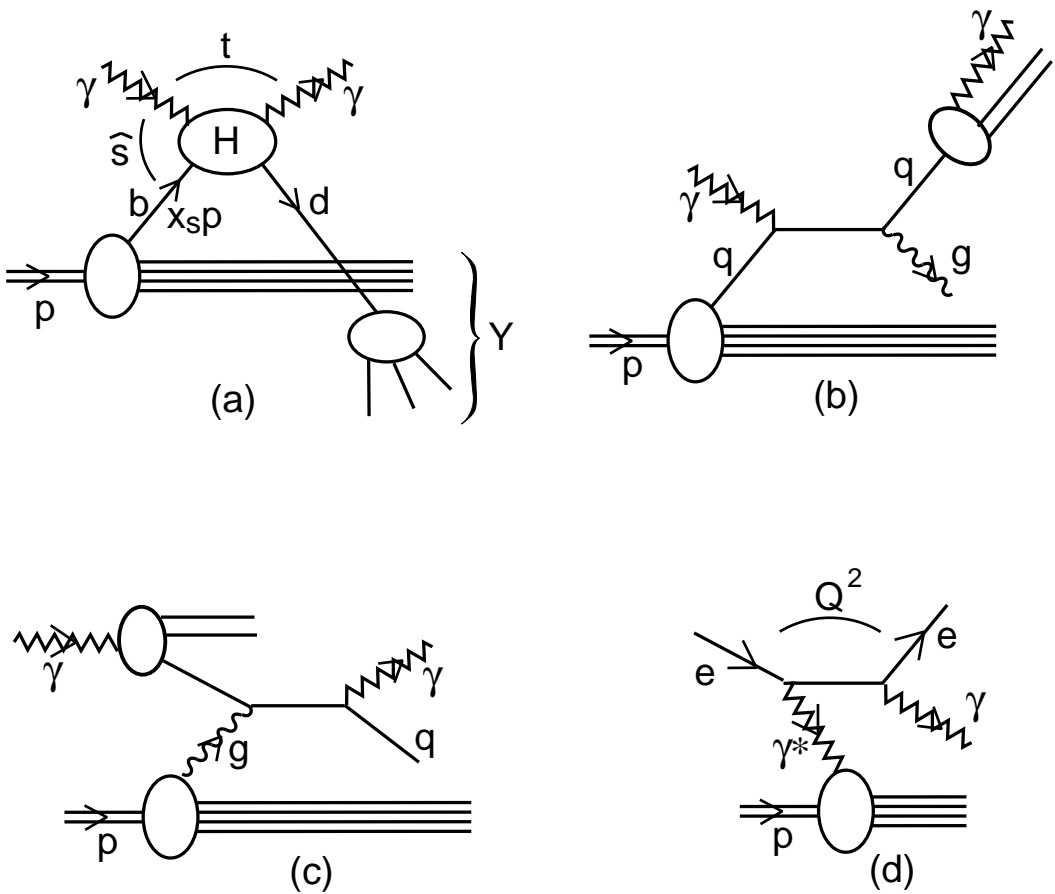}
\medskip
\caption{(a) Semi-exclusive photoproduction of photons. The hard process H
is calculable in PQCD. (b) Photons produced via parton fragmentation are
accompanied by comoving hadrons. (c) The resolved photon leaves remnant
hadrons in the beam direction. (d) The Bethe-Heitler process is not
enhanced at low momentum transfer $Q^2$ between the electrons.}
\label{kinem}
\end{figure}

We shall discuss the feasibility of measuring semi-exclusive photon
photoproduction, $\gamma p \to \gamma Y$ at HERA. Here the incoming and
outgoing photons {\em together} form an effective current which probes a parton
(quark or gluon) in the target with resolution $-t$ (Fig. 1a). This picture is
valid provided that the final photon emerges isolated from all hadrons in the
inclusive system $Y$, which is ensured by the kinematic requirement 
$W^2/M_Y^2 \gg
1$, where $W$ is the total $\gamma p$ center of mass energy, and 
$M_Y$ is the mass
of the hadronic system $Y$. Scattering on a single parton implies large
$-t$, typically of order $M_Y^2$. As we shall see, luminosity 
limitations impose
$-t \ll M_Y^2$ at HERA, implying small $x_s$ in \eq{xsdef}.

It is not {\em a priori} obvious how stringent the kinematic 
restrictions need to
be to ensure semi-exclusive dynamics. At moderate $W^2/M_Y^2$,
photons may be produced via parton fragmentation, following a hard
process like $\gamma q \to g q$ (Fig. 1b). Such a process will 
typically generate
hadrons comoving with the photon, implying $W^2/M_Y^2 \sim 1$. Depending
on the probability that the photon takes most of the momentum of the 
fragmenting
parton such processes can nevertheless be significant at moderate values of
$W^2/M_Y^2$. Based on the {\sc PYTHIA} \cite{Sjostrand:1994yb} and 
{\sc LUCIFER}
\cite{Ingelman:1987dt} Monte Carlo generators we conclude that fragmentation
processes are insignificant for $W^2/M_Y^2 \gsim 10$ and $-t \gsim 4\ \gev^2$.

Resolved photon contributions (Fig 1c) are analogously suppressed since
hadrons in the beam direction must carry little energy for $W^2/M_Y^2$ to be
large. This requires the beam parton to carry nearly all the
photon energy, and not to emit collinear bremsstrahlung prior to its hard
scattering. The pointlike photon component transfers the
full beam energy into the hard process and will thus dominate at large
$W^2/M_Y^2$ and $-t$.

We also estimate the contribution of the Bethe-Heitler process (Fig. 1d),
where the beam photon is virtual and the final photon is emitted
from the electron. This process is typically detected because of a
large angle recoil of the electron. We find that the Bethe-Heitler 
process is in
any case negligible for invariant momentum transfers $Q^2 \lsim 0.1\ \gev^2$
between the initial and final electrons.

HERA data \cite{Breitweg,Lee:1999sp} on photon production at large transverse
momentum have previously been compared with QCD calculations
\cite{duke,auranche1,auranche2} in terms of an isolation cut
\cite{Gordon:1998yt,Krawczyk:1998it,Frixione}, which imposes low
accompanying transverse energy in a cone around the photon direction. 
Such a cut
removes a large fraction of the photons produced by parton fragmentation, but
allows resolved photon contributions. The isolation procedure 
requires modelling
the non-perturbative fragmentation process. This is avoided for the 
semi-exclusive
cross section (\ref{gencross}), which depends only on standard 
structure functions
and perturbatively calculable quantities.

\begin{figure}[tb]
\center\leavevmode
\epsfxsize=10cm
\epsfbox{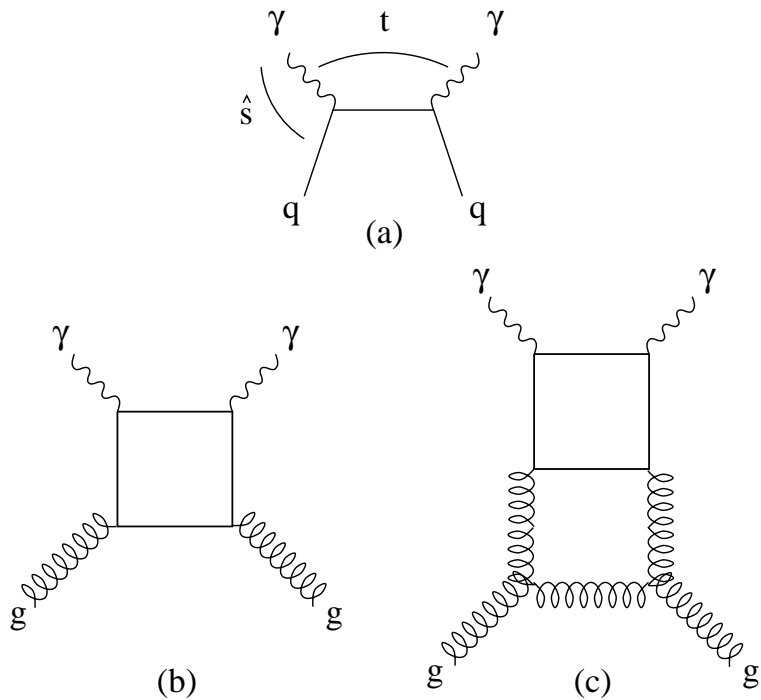}
\medskip
\caption{Compton scattering subprocesses: Low order scattering on a quark (a)
and on a gluon (b). At high $ \hat s/-t$, higher order amplitudes like (c)
dominate, due to gluon exchange in the $t$-channel.}
\label{graphs}
\end{figure}

Low order PQCD subprocess amplitudes which contribute to the hard vertex
$H$ of Fig. 1a are shown in Fig.~2. It is important to notice that the lowest
order contributions of Fig.~2a,b actually are suppressed in the Regge limit
$\hat s = x_s W^2 \to \infty$ at fixed (but large) $-t$, which is relevant for
semi-exclusive dynamics. A (dimensionless) amplitude involving the 
exchange of two
particles of spins $j_1$ and $j_2$ in the $t$-channel behaves in the 
Regge limit
as $\sim (\hat s/-t)^\alpha$, where $\alpha = j_1+j_2-1$. Hence the 
contribution
of Fig.~2b, with $j_1=j_2=1/2$, has $\alpha = 0$. The amplitude of
Fig.~2a also has $\alpha = 0$. On the other hand, the higher order amplitude of
Fig.~2c has $j_1=j_2=1$ and thus $\alpha = 1$. At even higher orders 
logarithmic
corrections from repeated gluon ladders are expected to build up hard Pomeron
exchange in the BFKL approximation, as has been studied extensively for the
present process
\cite{Ginzburg:1996vq,Ivanov:1998jw,Evanson:1999zb,Cox:1999kv}. 

It is interesting to explore at what value of $ \hat s/-t$ the higher order
diagrams begin to dominate.  We find that the BFKL amplitude is numerically
important compared to the lowest  order contributions in the whole 
semi-exclusive
range $ \hat s/-t \gsim 10$.  If the BFKL approximation is reliable (or the
amplitude of Fig.~\ref{graphs}c dominates)  at such moderate values of the
subprocess energy then the subprocess cross  section will be found to 
grow (or be
constant) in this whole range.  Alternatively, if the lowest order diagrams
dominate, the subprocess cross section will decrease in the lower range of
$\hat s/-t$. This would make it smaller than the BFKL estimate at high 
$\hat s/-t$.

\section{Subprocess Cross Sections} \label{secII}

The lowest order cross section for $\gamma q \to \gamma q$
(Fig.~\ref{graphs}a) is
\cite{Afanasev:1998ie}
\beq
\frac{d\sigma_{\rm LO}}{dt} (\gamma q \to \gamma q) =
\frac{2\pi e_q^4 \alpha^2}{ \hat s^2}
\left( \frac{ \hat s}{- \hat u} +  \frac{- \hat u}{ \hat s}\right) ,
\label{gamq}
\eeq
with $\alpha = e^2/4\pi$.
For the lowest order $\gamma g \to \gamma g$ process (Fig.~\ref{graphs}b)
we have analogously \cite{Doncheski:1992rf,Berger:1984yi},
\beq
\frac{d\sigma_{\rm LO}}{dt}(\gamma g \to \gamma g) =
2 \left( \sum_q e_q^2 \right)^2
\frac{\alpha^2 \as^2}{64 \pi  \hat s^2}
\left[|M_1( \hat s, t, \hat u)|^2 + |M_1'( \hat u,t, \hat s)|^2 +
|M_1'(t, \hat s, \hat u)|^2 + 20 \right] \,,
\label{gamg}
\eeq
where
\beqa
|M_1( \hat s, t,  \hat u)|^2 &=&
\left( 2 + 2 \frac{t -  \hat u}{ \hat s} \ln \frac{t}{ \hat u}
+\frac{ t^2 + \hat u^2}{ \hat s^2}
\left[ \ln^2 \frac{t}{ \hat u} + \pi^2 \right] \right)^2 ,
\nonumber \\
|M_1'( \hat u,t, \hat s)|^2 &=&
\left( 2 + 2 \frac{ \hat s - t}{ \hat u} \ln \frac{\hat s}{-t}
+\frac{ \hat s^2 + t^2}{ \hat u^2}
\ln^2 \frac{\hat s}{-t}\right)^2
+ 4 \pi^2 \left( \frac{\hat s^2 + t^2}{ \hat u^2}
\ln \frac{\hat s}{-t} + \frac{ \hat s -t}{ \hat u} \right)^2\;.
\label{Mexprs}
\eeqa
It is interesting to compare the size of the BFKL cross section to the
LO ones given above. The $\gamma g \to \gamma g$ BFKL cross section is
to leading logarithmic  accuracy given by
\cite{Ivanov:1998jw,Evanson:1999zb,Cox:1999kv}
\begin{eqnarray}
\frac{d\sigma_{\rm BFKL}}{dt}  (\gamma g \to \gamma g) &=&\frac{81}{16}
\frac{\alpha^2 \alpha_s^4}{576 t^2} \pi
\left( \sum_q e_q^2 \right)^2
\left[G\left(\frac{3 \as}{\pi} \ln \frac{ \hat s}{-t} \right) \right]^2 ,
\quad \textrm{with}
\nonumber \\
G(z) &=& \int_{-\infty}^\infty \frac{d \nu}{1+\nu^2}
                                 \frac{\nu^2}{\left(\nu^2+\frac{1}{4}\right)^2}
                                 \frac{\tanh(\pi \nu)}{\pi \nu}
                                 2(11+12\nu^2)
                                 \exp\left[z\chi(\nu)\right] \,,
\nonumber \\
\chi(\nu) &=& 2\psi(1)-2{\rm Re}\left[\psi\left(\half+i\nu\right)\right] \,.
\label{bfkleq}
\end{eqnarray}

\begin{figure}[htb]
\center \leavevmode
\epsfxsize=9cm
\epsfbox{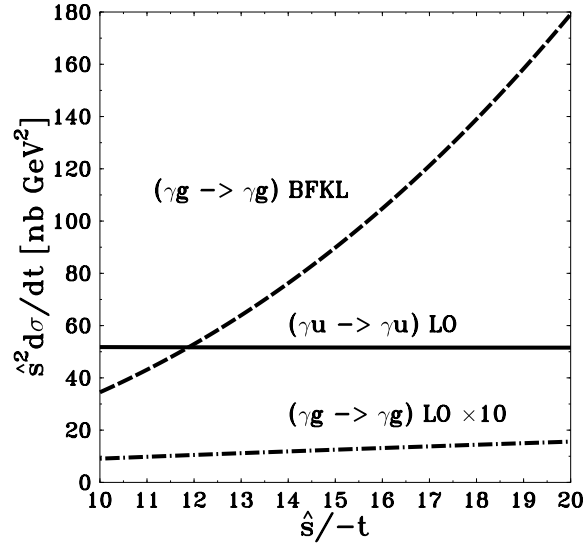}
\medskip
\caption{Lowest order subprocess cross sections ${ \hat s}^2 d\sigma/dt$ for
$\gamma u \to \gamma u$ (solid line) and $\gamma g \to \gamma g$ (times 10,
dot-dashed line) for $10 \leq  \hat s/-t \leq 20$.
The dashed line shows the BFKL cross section extrapolated to this region of
moderate subprocess energy, assuming $\as = 0.2$.
Three active flavors have been used in both the calculation of the
process $\gamma g \to \gamma g$ to LO and the BFKL contribution.}
\label{sigmas}
\end{figure}

In Fig.~\ref{sigmas} we show the dimensionless lowest order cross sections
${ \hat s}^2 d\sigma/dt$ for $\gamma u \to \gamma u$ (solid line) and
$\gamma g \to \gamma g$ (times 10, dot-dashed line). The small size of the
lowest order $\gamma g \to \gamma g$ cross section means that it will be
insignificant also in the physical process $\gamma N \to \gamma Y$, even though
the gluon distribution is larger than the quark one at low values of
$x_s$. The dashed line shows the BFKL cross section (\ref{bfkleq}) for $\gamma
g \to \gamma g$, based on iterating higher order diagrams like that in
Fig.~\ref{graphs}c. We used the fixed value $\as = 0.2$ indicated by BFKL fits
to HERA and Tevatron data \cite{Bartels:1996fs,ZEUS99,Cox:1999dw}. The BFKL
cross section is comparable to the lowest order one already at
$ \hat s/-t \simeq 10$, with their ratio growing roughly as $(\hat s/-t)^2$. It
should be stressed, however, that the BFKL analysis is expected to be reliable
only for $ \hat s/-t \gsim 100$
\cite{Cox:1999kv}, and that there may be sizeable corrections from NLO
corrections \cite{Fadin:1998py}.

\section{Semi-exclusive photon electroproduction at HERA}
\label{sec2}

The $ep \to e \gamma Y$ electroproduction cross section is in
the semi-exclusive limit $\lqcd^2 \ll -t, M_Y^2 \ll W^2$ related to the
subprocesses  through
\beq
\frac{d\sigma}{dy\,dt\,dx_s} (ep \to e \gamma Y) =
f_{\gamma/e}(y,Q_{\rm max}^2) \sum_{i\in q,g} f_{i/p}(x_s,-t)
\frac{d\sigma}{dt}
(\gamma i \to \gamma i) \,.
\label{epsigma}
\eeq
Here $y$ is the fraction of momentum that the photon, of virtuality
$\leq Q_{\rm max}^2$, takes of the electron beam.
The Weizs\"acker-Williams function $f_{\gamma/e}(y,Q^2_{\rm max})$ is given in
\cite{Frixione:1993yw},
\begin{equation}
f_{\gamma/e}(y,Q^2_{\rm max})
= \frac{\alpha}{2 \pi}
\left[ \frac{(1 + (1-y)^2)}{y} \ln \frac{Q^2_{\rm max}(1-y)}{m_e^2 y^2}
+2 m_e^2 y \left( \frac{1}{Q^2_{\rm max}} - \frac{1-y}{m_e^2
y^2}\right)\right].
\end{equation}
In our kinematical range the Weizs\"acker-Williams scale is identical
to the upper limit of the momentum transfer $Q_{max}^2$ between the 
electrons, as
the hard scale $-t$, which characterizes the production process, is much larger
than $Q^2_{max}$. A detailed discussion of the choice of the
Weizs\"acker-Williams scale can be found in Refs.
\cite{Frixione:1993yw,Frixione:1997ks}.
For the parton distributions $f_{i/p}(x_s,-t)$ we use the GRV94 LO
parameterizations \cite{Gluck:1995uf}.

In this section we first estimate the kinematic region in which the
semi-exclusive production mechanism dominates, and then evaluate the physical
cross section in this region.

\subsection{Background from Fragmentation and Decay in Photoproduction}

Large transverse momentum photons can be produced through quark and
gluon fragmentation $q,g \to \gamma + X$, following standard hard
scattering processes such as photon gluon fusion $\gamma g \to q\bar q$ and
gluon bremsstrahlung $\gamma q \to q g$.
A second source of contributions are photons from hadronic decays
like $\pi^0 \to \gamma \gamma$.
In the following we call both background contributions `fragmentation'.
Such processes typically give hadrons in the photon direction, and are thus
suppressed in the limit where $M_Y^2 \ll W^2$.

We have used the Monte Carlo event generators {\sc PYTHIA}
\cite{Sjostrand:1994yb} and {\sc LUCIFER} \cite{Ingelman:1987dt} to estimate
the range of $W^2/M_Y^2$ in which semi-exclusive production dominates over
fragmentation.  These event generators include the direct
(semi-exclusive) $\gamma q \to \gamma
q$ cross section only at lowest order.
As we observed in the introduction, the higher order contribution of Fig.~2c
(and its possible BFKL enhancement) actually dominates the lowest order
process (Fig.~2a) at high $\hat s/-t$. Hence comparing the fragmentation
background to only the LO semi-exclusive cross-section gives a conservative
estimate of the kinematic region where semi-exclusive dynamics dominates. It
should also be kept in mind that the fragmentation contribution can be reduced
with the help of photon isolation cuts
\cite{Breitweg,Lee:1999sp,Gordon:1998yt,Krawczyk:1998it,Frixione}.

We evaluate the contributions from the direct process
$\gamma q \to \gamma q$ and that from the background $\gamma g \to q\bar q$
and $\gamma q \to q g$ processes separately.
If the final state contains several photons, we choose the one with largest
energy $E_\gamma^{CM}$ in the CM frame of the photon and the nucleon.
The invariant mass $M_Y^2$ of the remaining particles is then given by
$M_Y^2 = W^2 -2 W E_\gamma^{CM}$.

\begin{figure}[tb]
\centerline{\psfig{figure=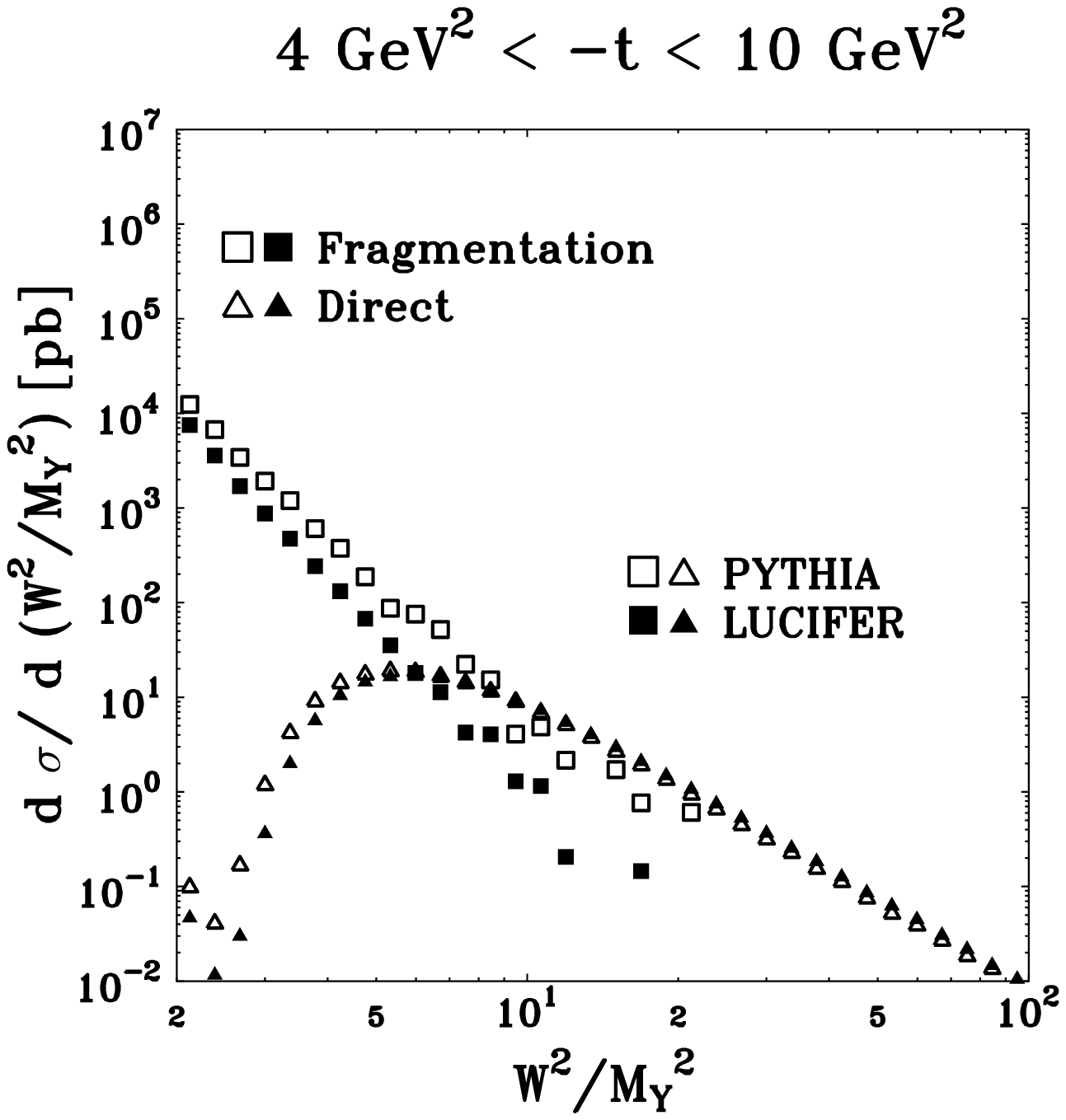,width=9cm}
              \psfig{figure=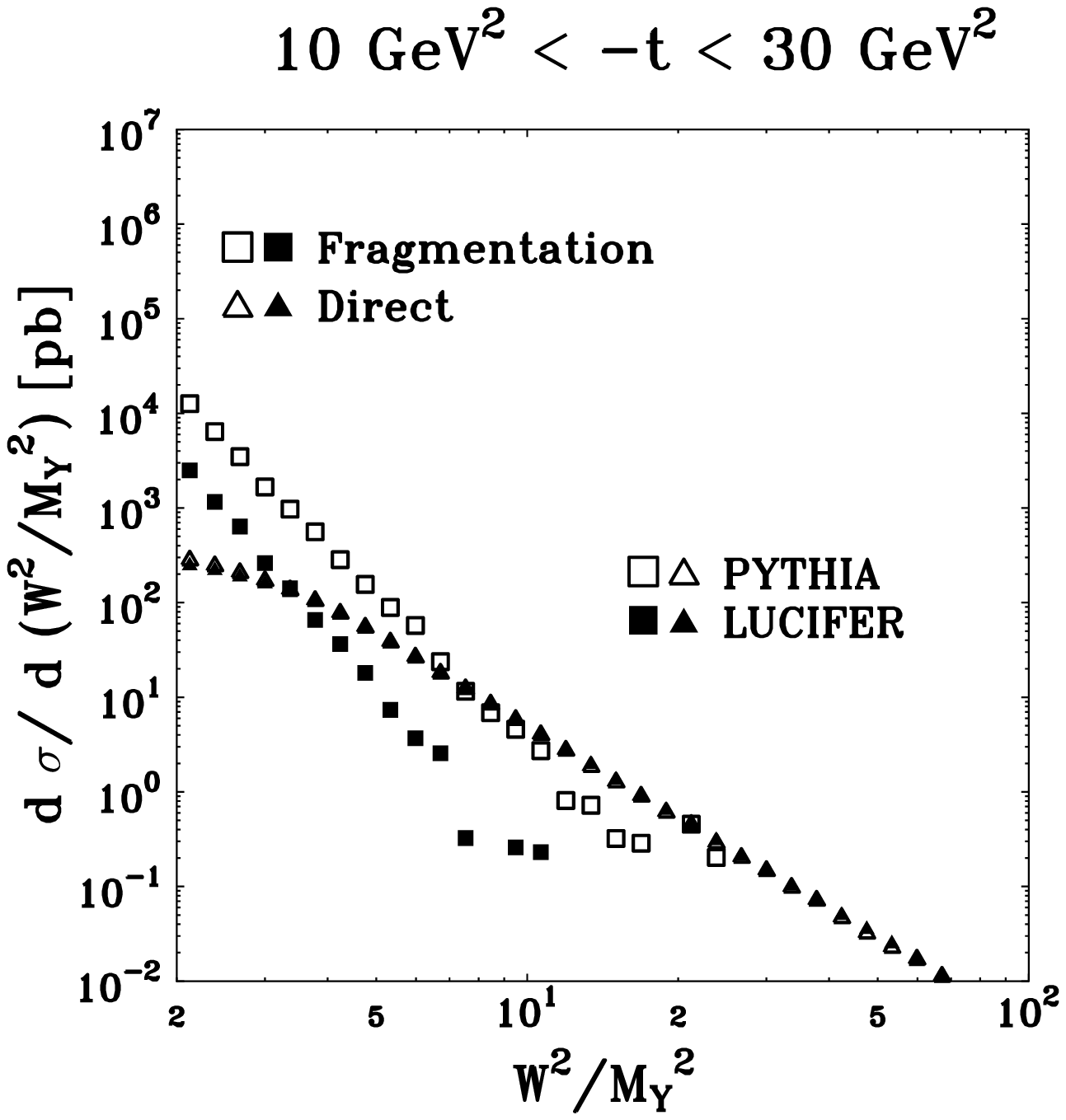,width=9cm}}
\caption{
Inclusive cross sections $d\sigma(\gamma p\to\gamma Y)/d(W^2/M_Y^2)$ for
the photon with the largest $E_\gamma^{CM}$ per event at $W = 200 \;{\rm
GeV}$. The $t$-range is
$4 \; {\rm GeV}^2 < -t < 10 \;{\rm GeV}^2$ (left), and
$10\; {\rm GeV}^2 < -t < 30 \;{\rm GeV}^2$ (right).
The triangles show the direct
$(\gamma q\to \gamma q)$ contribution of Eq. (3) and the
squares show photons generated by fragmentation.  The open squares and
triangles are the result of the {\sc PYTHIA} program and the filled squares
and triangles are from the {\sc LUCIFER}  run.
An additional cut  $.001 <x_s < .7$ has been applied in  both figures.
Higher order parton showering and resolved photon contributions are not
included.}
\label{mc}
\end{figure}

In Fig.~\ref{mc} we show the {\sc PYTHIA} and {\sc LUCIFER} results for
$W=200 \ \gev$ and two ranges of photon transverse momentum,
$4\ \gev^2 < -t < 10\ \gev^2$ and $10\ \gev^2 <-t < 30\ \gev^2$.
It may be seen that the semi-exclusive process $\gamma q \to \gamma
q$ starts to
dominate for $W^2/M_Y^2 \gsim 10$.
We have checked that the same conclusion holds for $W=80\ \gev$.
Note that for $M_Y^2 \gg -t$, the subprocess variable $\hat s/-t$ of
Fig.~\ref{sigmas} is in the semi-exclusive limit simply related to
$W^2/M_Y^2$,

\beq
\frac{W^2}{M_Y^2} = \frac{ \hat s}{-t}\left(1-\frac{t}{M_Y^2} \right) \simeq
\frac{ \hat s}{-t} \,.
\label{kinrel}
\eeq

\subsection{Background from the Bethe-Heitler Process in Electroproduction}

\begin{figure}[htb]
\center\leavevmode
\epsfxsize=14cm
\epsfbox{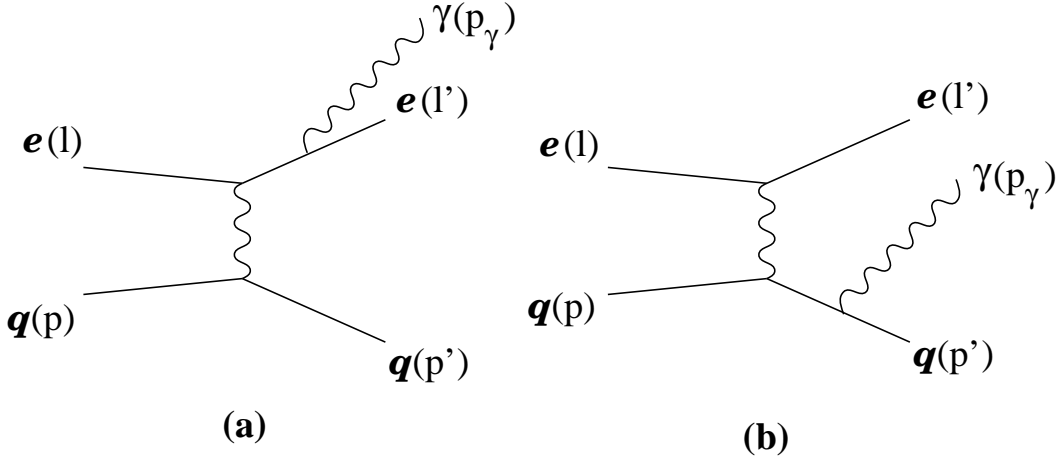}
\medskip
\caption{Sample diagrams for photon production via the Bethe-Heitler (BH)
process (a) and Virtual Compton Scattering (VCS) (b).}
\label{bh}
\end{figure}

When the incoming photon is virtual the final photon may
be radiated off the electron in the Bethe-Heitler (BH) process of
Fig.~\ref{bh} a. We wish to determine the maximum value of $Q^2$ for 
which the BH
cross section can be neglected compared to that of photon emission 
from the quark,
\ie, the Virtual Compton Scattering (VCS) process of Fig.~\ref{bh} b.

At finite $Q^2$ the lowest order $ep \to e\gamma Y$ cross section can be
written
\beqa
\frac{d\sigma_{LO}(ep \to e\gamma Y)}{dx_s\,dW^2\,dQ^2\, dt\,d\phi}
&= &\frac{1}{2(4\pi)^4 x_s W_{ep}^4 (\hat{s} + Q^2)} \sum_q
f_{q/p}(x_s,-t)\overline{|{\cal M}_q|^2} \,,
\quad \textrm{with} \label{epdiff}
\\ \vspace{2mm}
\overline{|{\cal M}_q|^2} & = &
\overline{|{\cal M}_q^{VCS}|^2} + \overline{|{\cal M}_q^{BH}|^2} +
\overline{2 {\rm Re}\left({\cal M}_q^{BH} {{\cal
M}_q^{VCS}}^* \right)}\,.
\nonumber
\eeqa
Here $Q^2 = - q^2 = -(l-l')^2$ and
$\hat{s} = (q + p)^2 = x_s W_{\gamma p}^2$ in the notation of Fig.~\ref{bh}.
$\phi$ is the angle between $\vec q \times \vec p_{\gamma}$ and
$\vec\ell \times \vec\ell'$ with the three-vectors given in the $CM$ frame of
the virtual photon and the proton, while $W_{ep} \approx 300 \ \gev$ is the
$ep\ CM$ energy. We also define
\beq
\hat{s}_l = (l'+p_\gamma)^2 \,, \quad
\hat{u}_l = (\l - p_\gamma)^2
\label{varlept}
\eeq
which can be obtained from $\hat{s}$ and $\hat{u}$ by replacing the quark
momentum $p(p')$ with the lepton momentum $l(l')$.

The squared VCS and BH matrix elements can now be written as
\beqa
\overline{|{\cal M}_q^{VCS}|^2} &=&
\frac{-4 (4\pi\alpha)^3 e_q^4}{Q^2  \hat{s}  \hat{u}}F \,,
\label{MVCS}  \\
\overline{|{\cal M}_q^{BH}|^2} &=&
\frac{4 (4\pi\alpha)^3 e_q^2}{t \hat{s}_l \hat{u}_l}F \,,
\label{MBH}
\eeqa
with the common factor $F$ given by
\beq
F = { \hat s}^2 + ( \hat s+t)^2 + \hat{u}_l^2
+ (2W_{eq}^2 + \hat{u}_l)^2 - 2 \hat s (2W_{eq}^2 + \hat{u}_l)
- 2t (W_{eq}^2 + \hat{u}_l) + 2Q^2 (2 \hat s + t - 3W_{eq}^2 - 2\hat{u}_l)
+ 3Q^4 \,,
\eeq
and $W_{eq}^2=(\l+p)^2=x_s W_{ep}^2$.

\newpage
\begin{eqnarray}
\overline{2 {\rm Re}\left({\cal M}_q^{BH} {{\cal M}_q^{VCS}}^* \right)} & = &
\frac{8 (4\pi\alpha)^3 e_q^3}{Q^2 t}
\Bigg\{ Q^2
\left[-\frac{1}{2}( \hat s - \hat u)
\left(\frac{1}{\hat{s}_l} - \frac{1}{\hat{u}_l}\right)
+ \frac{1}{\hat{u}_l}(l \cdot p + l  \cdot p') +
    \frac{1}{\hat{s}_l}(l'\cdot p + l' \cdot p')\right]
\nonumber \\
&& \qquad \quad \; + t \left[\frac{1}{2}(\hat{s}_l - \hat{u}_l)
\left(\frac{1}{  \hat s}-\frac{1}{  \hat u}\right)
- \frac{1}{ \hat u}(l \cdot p + l' \cdot p )
- \frac{1}{ \hat s}(l \cdot p'+ l' \cdot p')\right]
\nonumber \\
&& \qquad \quad \;
-8 (l\cdot p \;l'\cdot p' + l\cdot p'\; l'\cdot p)
\left(
\frac{l \cdot p }{ \hat u \hat{u}_l} +
\frac{l \cdot p'}{ \hat s \hat{u}_l} +
\frac{l'\cdot p }{ \hat u \hat{s}_l} +
\frac{l'\cdot p'}{ \hat s \hat{s}_l}\right)
\nonumber \\
&& \qquad \quad \;
+l\cdot p \left[
\frac{ \hat s \hat{s}_l}{ \hat u  \hat{u}_l} - 3
+ \frac{2}{ \hat u \hat{u}_l} \left(  \hat s l'\cdot p + \hat{s}_l l\cdot p'
-( \hat u + \hat{u}_l) l'\cdot p' \right)\right]
\nonumber \\
&& \qquad \quad \;
+l \cdot p' \left[ \frac{ \hat u \hat{s}_l}{ \hat s \hat{u}_l} - 3
+ \frac{2}{ \hat s \hat{u}_l }\left(-\hat{s}_l l\cdot p -  \hat u l'\cdot p'
+ ( \hat s+ \hat{u}_l) l'\cdot p \right)\right]
\nonumber \\
&& \qquad \quad \;
+l' \cdot p \left[ \frac{ \hat s \hat{u}_l}{ \hat u \hat{s}_l} -3
+ \frac{2}{ \hat u \hat{s}_l}\left( - \hat s l \cdot p - \hat{u}_l l'\cdot p'
+ ( \hat u + \hat{s}_l) l \cdot p' \right) \right]
\nonumber \\
&& \qquad \quad \;
+l'\cdot p' \left[
\frac{ \hat u \hat{u}_l}{ \hat s \hat{s}_l} - 3 + \frac{2}{ \hat s \hat{s}_l}
\left( \hat u l\cdot p' + \hat{u}_l l'\cdot p
- ( \hat s + \hat{s}_l) l\cdot p \right)\right]
\Bigg\} . \label{interf}
\end{eqnarray}

The results presented in Eqs. (\ref{MVCS}) - (\ref{interf}) have previously
been computed, \eg, in Ref. \cite{Brodsky:1972yx}, but with our particular
choice of kinematical variables we obtained more compact expressions.

\begin{figure}[htb]
\center\leavevmode
\epsfxsize=12cm
\epsfbox{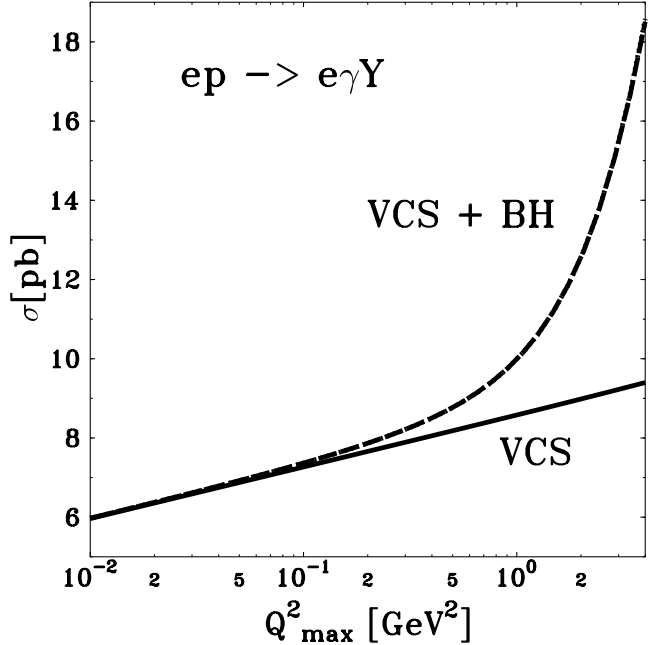}
\medskip
\caption{Dependence of the total $ep\to e\gamma Y$ cross section on the
maximal momentum transfer $Q^2_{\rm max}$. The solid line shows the
contribution of Virtual Compton Scattering (VCS, Fig. 5b) alone, while the
dashed line includes also the Bethe-Heitler (BH, Fig. 5a) contribution and the
interference term. The kinematic region is defined in the text.}
\label{background}
\end{figure}
%
%
%
In Fig.~\ref{background} we compare the VCS to the total VCS+BH cross section
for $ep\to e\gamma Y$ as a function of $Q_{\rm max}^2$, the maximum momentum
transfer between the electrons. The differential cross section 
(\ref{epdiff}) was
integrated over the ranges $x_s \in [0.01,0.7]$, $W \in [40\ \gev,160\ \gev]$,
$\phi \in [0,2\pi]$, $-t \in [4\ \gev^2,10\ \gev^2]$, and
$Q^2 \in [m_e^2, Q_{\rm max}^2]$\footnote{In order to ensure that the electron
propagators are far off-shell we also required $\hat{s}_l,-\hat{u}_l > -t_{\rm
min} = 4 \;{\rm GeV}^2$. This cut is, however, irrelevant for $Q^2
\lsim 1\ \gev^2$.}.

\subsection{The Semi-Exclusive Cross Section}

\begin{figure}[htb]
\center\leavevmode
\epsfxsize=8cm
\epsfbox{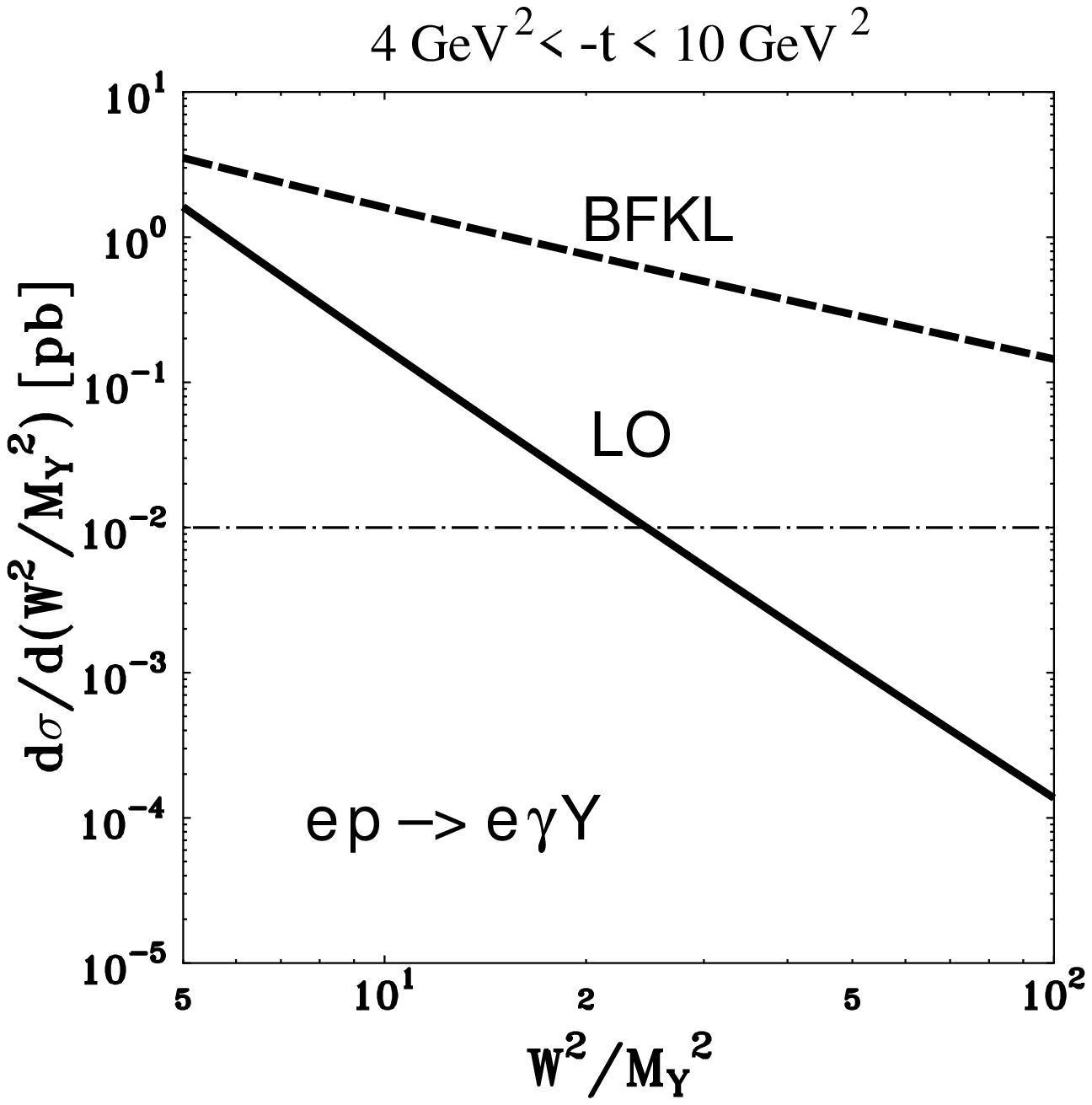}
\epsfxsize=8cm
\epsfbox{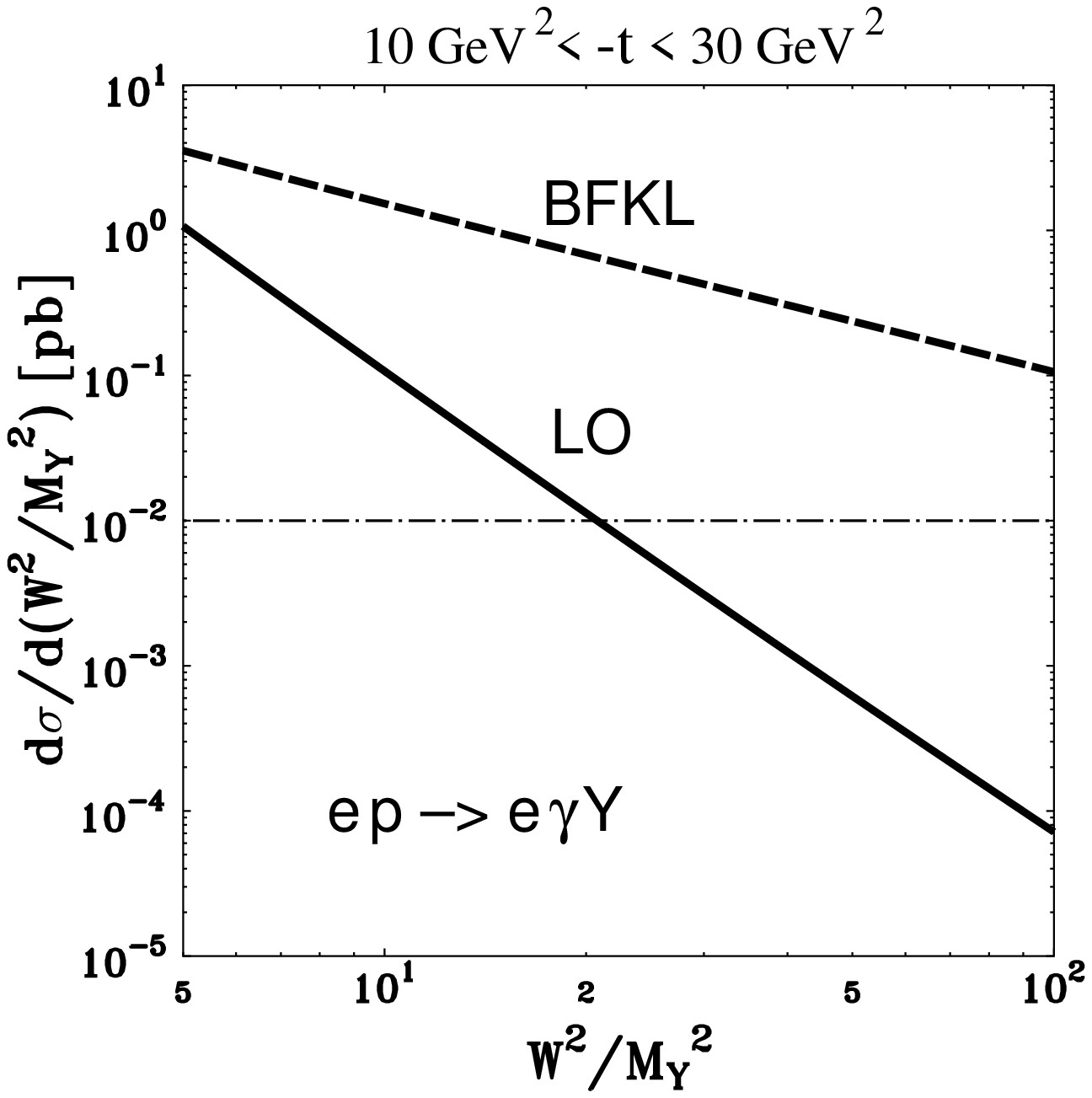}
\medskip
\caption{
The cross section $d\sigma(ep\to e\gamma Y)/d(W^2/M_Y^2)$ of Eq.~(16) shown for
two ranges of $t$: $4\ \gev^2$$<-t$$<10\ \gev^2$ (left) and
$10\ \gev^2<$$-t$$<30\ \gev^2$ (right).
The solid line shows the contribution of the LO process of Eq.~(3) (Fig.~2a).
The dashed line shows that of the BFKL process of Eq.~(17) (Fig.~2c plus gluon
ladder iterations). The horizontal dash-dotted line gives the 1
event level per unit of $W^2/M_Y^2$, for a HERA luminosity of 100
events/pb.}
\label{fig7}
\end{figure}

We have found that the process $ep \to e \gamma Y$ can be used to study
semi-exclusive photon production at HERA in the kinematic range
$W^2/M_Y^2 \gsim
10$, $Q^2 \lsim 0.1\ \gev^2$ and $-t \gsim 4\ \gev^2$.
In Fig.~\ref{fig7} we show the cross section
\beq
\frac{d\sigma (ep \to e \gamma Y)}{d(W^2/M_Y^2)} = \int_{0.25}^{0.75} dy
   f_{\gamma/e}(y,Q_{\rm max}^2= 0.1\ \gev^2) \int_{|t|_{\rm
min}}^{|t|_{\rm max}} dt
\frac{-t(1-x_s)^2}{W^2} \sum_{i\in q,g} f_{i/p}(x_s,-t)
\frac{d\sigma}{dt} (\gamma i \to \gamma i)
\label{sepsigma}
\eeq
for two ranges of momentum transfer,
$4\ \gev^2<-t<10\ \gev^2$ and $10\ \gev^2<-t<30\ \gev^2$.
The full curve shows the contribution of the LO process of \eq{gamq}
(Fig.~\ref{graphs}a).  The dashed curve shows the contribution of the BFKL
subprocess of \eq{bfkleq} (Fig.~\ref{graphs}c, plus gluon ladder iterations),
together with the corresponding $\gamma q \to \gamma q$ sea quark BFKL
contribution \cite{Evanson:1999zb},
\begin{equation}
\frac{d\sigma_{\rm BFKL}}{dx_s d t} =
\left[g(x_s,-t) + \frac{16}{81} \Sigma(x_s,-t)\right]
\frac{d \sigma_{\rm BFKL}}{d t} (\gamma g \to \gamma g)\;,
\label{bfklevanson}
\end{equation}
%
%
%
%
where $\Sigma$ denotes the quark singlet distribution
$\Sigma = \sum_q [f_{q/p} + \bar f_{q/p}]$.
We use $\alpha_s = 0.2$ in the BFKL cross section and assume three (four)
active flavors in the low (high) $|t|$-range. The horizontal dash-dotted line
indicates the 1 event level per unit of $W^2/M_Y^2$, given a nominal HERA
luminosity of 100 events/pb.

We should emphasize
that the BFKL approximation may not be reliable for
$W^2/M_Y^2 \lsim 100$ \cite{Cox:1999kv}, hence the dashed curve is an
extrapolation. It is nevertheless interesting to observe that this
extrapolation dominates the LO $q\gamma \to q\gamma$ contribution over the
whole semi-exclusive range $W^2/M_Y^2 \gsim 10$. This conclusion is insensitive
to the $t$-range and also to the range of $y$ (not shown).

\section{Conclusions}

Semi-exclusive processes $A+B\to C+Y$ provide a new tool for investigating
hadron structure. Effective currents formed by the $A\overline C$ system
generalize the virtual photon probe familiar from DIS and
can carry charge, flavor, baryon and other quantum numbers
\cite{Brodsky:1998sr}. Before this tool can be put to use, at least two
questions need to be answered:

\noindent
(i) {\em How stringent limits $\lqcd^2 \ll -t,M_Y^2 \ll W^2$ must be imposed in
order for the semi-exclusive production mechanism to dominate?}\\
(ii) {\em Can the hard $A\overline C$ vertex be reliably computed using PQCD?}

In this paper we studied the process $\gamma p \to \gamma Y$, which is
especially simple in the sense that both particles $A$ and $C$ have a 
point like
component. Based on simulations with the PYTHIA and LUCIFER event generators we
concluded that semi-exclusive dynamics should dominate for $W^2/M_Y^2 \gsim
10$ and $-t \gsim 4\ \gev^2$. Photon emission from the electron (the
Bethe-Heitler process) is insignificant for incoming photon virtualities $Q^2
\lsim 0.1\ \gev^2$, and can be further suppressed with angular cuts. The
semi-exclusive cross section should be measurable at HERA, assuming
that the subprocess cross section $\hat\sigma(\gamma q \to\gamma q)$ is not
smaller than its lowest order (LO) PQCD approximation.

Point (ii) above is non-trivial, since the semi-exclusive kinematics implies
a high energy (Regge) limit for the subprocess, $\hat s/-t \gg 1$. Little is
known about the importance of higher order (HO) PQCD corrections in this limit.
In the process under study the situation is particularly intriguing since the
LO subprocess diagrams shown in Fig.~\ref{graphs}a,b correspond to
$q\bar q$ exchange in
the $t$-channel. At high subenergies $\hat s$ they are therefore
suppressed by a
factor $1/{\hat s}^2$ in the cross section compared to the \order{\as^4} gluon
exchange contribution of Fig.~\ref{graphs}c. The latter is, on the
other hand, just the
first term in the series of gluon ladder diagrams which is supposed to build up
the BFKL Pomeron in this process
\cite{Ivanov:1998jw,Evanson:1999zb,Cox:1999kv}.

It is not clear from which value of $\hat s/-t$ HO contributions like
Fig.~\ref{graphs}c start to dominate the LO processes of
Fig.~\ref{graphs}a,b. The BFKL
approximation has been assumed to be relevant for $\hat s/-t \gsim 100$
\cite{Cox:1999kv}. Extrapolating the BFKL cross section to lower energies we
found (Fig.~\ref{fig7}) that it would in fact dominate the LO cross
section in the whole
range of semi-exclusive dynamics, $\hat s/-t \gsim 10$. The ratio $\sigma_{HO}/
\sigma_{LO}$ behaves approximately like $(W^2/M_Y^2)^2$, closely
reflecting the $\hat s/-t$ dependence
of the respective subprocesses, \cf \eq{kinrel}.
A measurement of the $(W^2/M_Y^2)^2$ dependence of the cross section will thus
directly determine the nature of the dominant $t$-channel exchange.

We conclude that the large HERA energy in principle allows accessing the
semi-exclusive kinematic region, with its double hierarchy of large scales. The
limiting factor will be the luminosity. If the cross section is approximately 
given by the lowest order contribution in Fig.~7 then only a restricted range of
$(W^2/M_Y^2)^2$ can be studied. Since the higher order two gluon exchange
(BFKL) contributions fall off more slowly with $(W^2/M_Y^2)^2$ they will
eventually dominate. If their normalization is even close to that
indicated by the BFKL extrapolation of Fig.~7 there should be a rich
semi-exclusive phenomenology at HERA, not only for photon but also for meson
($\pi,\ \rho,\ J/\psi$) production. Further work is needed to estimate the
feasibility of measuring charge exchange processes and the three gluon
(Odderon) contribution to $\pi^0$ production.

\noindent
{\bf Acknowledgements.} We are grateful for helpful discussions with Stan
Brodsky and Markus Diehl. This work is supported in part by the EU/TMR
contract EBR FMRX-CT96-0008.

\end{document}